\documentclass[reprint,aps,prb,reprintnumbers,amsmath,amssymb,superscriptaddress]{revtex4-2}

\usepackage{graphicx}% Include figure files
\usepackage{dcolumn}% Align table columns on decimal point
\usepackage{bm}% bold math
\usepackage{upgreek}

\usepackage{epstopdf} 
\newcommand{\D}{\mathrm{D}}
\newcommand{\G}{\mathrm{G}}

\usepackage[multiple]{footmisc}

\begin{document}

\title{Cryogenic temperature dependence and hysteresis of surface-trap-induced gate leakage in GaN high-electron-mobility transistors}

\author{Ching-Yang Pan}
\affiliation{Department of Physics, National Taiwan Normal University, Taipei 116, Taiwan}
\author{Shi-Kai Lin}
\affiliation{Department of Physics, National Taiwan Normal University, Taipei 116, Taiwan}
\author{Yu-An Chen}
\affiliation{Department of Physics, National Taiwan Normal University, Taipei 116, Taiwan}
\author{Pei-hsun Jiang}
\email{pjiang@ntnu.edu.tw}
\affiliation{Department of Physics, National Taiwan Normal University, Taipei 116, Taiwan}

\begin{abstract}
This work provides a detailed mapping of various mechanisms of surface-trap-induced gate leakage in GaN HEMTs across a temperature range from room to cryogenic levels. Two-dimensional variable-range hopping is observed at small gate bias. Under higher reverse gate bias, the leakage is dominated by the Poole--Frenkel emission above 220 K, but gradually transitions to the trap-assisted tunneling below 220 K owing to the frozen-trap effect. The trap barrier height extracted from the gate leakage current under the upward gate sweep is 0.65 V, which is 12\% higher than that from the downward sweep. The gate leakage current as a function of the gate bias exhibits clockwise hysteresis loops above 220 K but counterclockwise ones below 220 K. This remarkable opposite hysteresis phenomenon is thoroughly explained by the trap mechanisms. 
\end{abstract}

\maketitle

\section{Introduction}
The demand for electronic components in ultralow temperature environments such as aerospace \cite{Bautista2001,Schleeh2013,Gui2020}, quantum computing \cite{Cha2020,HooTeo2021,Xie2021,Hornibrook2015}, and superconducting systems \cite{Rajashekara2013,Yang2012} continues to rise. Traditional field-effect transistors suffer from carrier freeze out, resulting in performance degradation at low temperatures. In contrast, gallium nitride high-electron-mobility transistors (GaN HEMTs), which typically exhibit high values of sheet carrier density (approximately $10^{13}$ cm$^{-2}$) and mobility (1000--2000 cm$^2$/(V s)) \cite{Bi2017,Roccaforte2018,Hospodkova2023,Pandey2020,Lee2016}, show no sign of carrier freeze out \cite{Gui2020a} and demonstrate superior performance at low temperatures, including low on-state resistance, low threshold voltage, low switching loss, high switching speed, improved subthreshold swing, and reduced thermal noise \cite{Wei2022,Wang2017,Yang2011,Jeon2021,Gokden2004, Fang2019}. However, defects in GaN HEMTs due to imperfect crystal arrangement can serve as carrier traps that hamper the device reliability, particularly resulting in trap-assisted gate leakage. The reverse gate leakage current, which can reduce the breakdown voltage and increase the off-state power dissipation \cite{Xu2018,Arulkumaran2003}, includes not only the bulk leakage vertically through the insulating layer \cite{Mitrofanov2004,Hsu2001,Hasegawa2003,Jiang2023}, but also the lateral leakage on the surface between the gate and the source and drain electrodes \cite{Zheng2014,Tan2006,Chen2014,Liu2011,Zheng2015}. Surface leakage increases significantly as devices downscale \cite{Kotani2007,Vetury2001}, and possesses a temperature dependence more drastic than the bulk leakage \cite{Tan2006}. To date, investigations of the temperature dependence of the gate leakage have primarily focused on temperatures above room temperature. Cryogenic performances of GaN HEMTs have started to attract greater attention recently \cite{Wei2022}; however, a thorough study of the cryogenic temperature dependence of their surface gate leakage is still lacking. 

On the other hand, the traps in GaN are thought to be the probable dominant mechanism of the kink effect \cite{Wang2011,Ma2014,Singh2018,Grupen2019}, a hysteresis in the drain output characteristics observed at low drain bias. The kink effect is usually attributed to the impact ionization in the conducting channel, or to the trapped states from the surface or from the bulk interface, and exhibits strong temperature dependence with apparent enhancement of the effect at lower temperatures \cite{Lin2005,Brar2002,Nazir2022,Cuerdo2009,Bisi2020}. This indicates a time-constant--related mechanism involved in the carrier transport in the kink effect, but identifying the underlying cause requires additional experimental techniques to probe transient time-dependent phenomena. While the physical origin of kink in the drain current of GaN HEMTs has been extensively studied, it remains not fully clarified \cite{Mao2021}, and studies examining hysteresis behaviors in trap-induced gate leakage are rare, with only a few isolated works briefly mentioning them in the literature \cite{Hult2022}. Not until very recently did the hysteresis of reverse leakage current in GaN Schottky-barrier diodes receive significant attention and prompt studies on its underlying mechanism \cite{Orfao2024, Pena2024}. The hysteresis phenomenon specifically identified the bias-induced occupancy of the energy states originating leakage-current processes. This insight can help develop a more comprehensive physical model for simulating the dynamics in the trap energy band. Therefore, investigations of GaN HEMTs on possible hystereses of the leakage current are also needed to promote better understanding of the trapping mechanisms.  

In this work, the gate leakage of GaN HEMTs is carefully explored at temperatures from 300 to 1.5 K. The leakage mechanism under positive gate bias is primarily Schottky thermionic emission (TE), but is found to be changed to \textit{surface} leakage under lower and negative gate biases, which involves two-dimensional variable-range hopping (2D-VRH) near zero gate bias, and frozen-trap stimulated transition around 220 K from the Poole--Frenkel emission (PFE) to the trap-assisted tunneling (TAT) under negative gate bias. This transition leads to remarkable opposite hystereses of the leakage current, pointing to a key difference between the dynamics of PFE and TAT during gate sweeps. The investigation and interpretation of the cryogenic hysteresis effect is the main focus of this work. The temperature dependence and the hysteresis behavior of the surface gate leakage can be crucial for development and applications of cryogenic devices.

\section{Device Structure and Fabrication}

\begin{figure}[!t]
\centering
\includegraphics[width=2.5in]{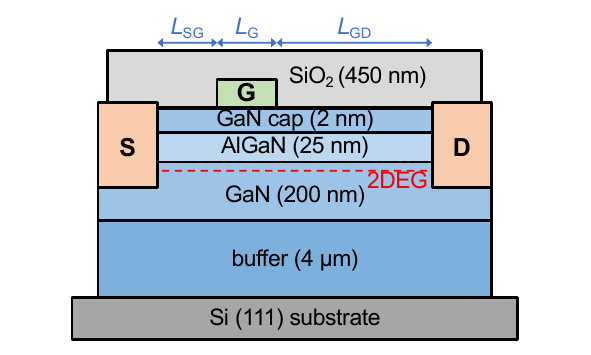}
\caption{\label{fig:device}
Schematic drawing (not to scale) of the GaN HEMT.
}
\end{figure}

The fabrication process of the GaN HEMTs is illustrated in Fig.~\ref{fig:device}. A 4-$\upmu$m-thick GaN/AlGaN superlattice consisting of periods of GaN/AlGaN layer pairs was deposited via metal--organic chemical vapor deposition (MOCVD) as a buffer layer on a Si(111) substrate to mitigate the dislocation density \cite{Sun2023,Tajalli2020}. Subsequently, a heterojunction of GaN (200 nm)/Al$_{0.2}$Ga$_{0.8}$N (25 nm) was grown using MOCVD to create a two-dimensional electron gas (2DEG) channel at the interface, and was then capped with a 2-nm GaN. Afterwards, source and drain Ti/Al/Ti/Au electrodes were formed and then annealed at 900 °C to ensure Ohmic contacts, followed by the formation of gate Ni/Au/Ti electrodes.  Finally, a passivation layer of SiO$_2$ (450 nm) was deposited using plasma-enhanced chemical vapor deposition. 

Electrical measurements at room and cryogenic temperatures were performed on several GaN HEMTs with a gate width of 100 $\upmu$m, various gate lengths ($L_\mathrm{G}$) from 3 to 5 $\upmu$m, source-to-gate lengths ($L_\mathrm{SG}$) from 3 to 5 $\upmu$m, and gate-to-drain lengths ($L_\mathrm{GD}$) from 20 to 30 $\upmu$m. Similar behaviors have been observed from all of them. The data presented in this letter were collected from a representative sample with $L_\mathrm{G}=5$ $\upmu$m, $L_\mathrm{SG}=3$ $\upmu$m, and $L_\mathrm{GD}=30$ $\upmu$m. Low-temperature measurements were conducted with the GaN HEMTs mounted in a cryogenic system. Connections to the electrodes were made via wire bonding. 

\section{Output and Transfer Characteristics}

\begin{figure}[!t]
\centering
\includegraphics[width=3.4in]{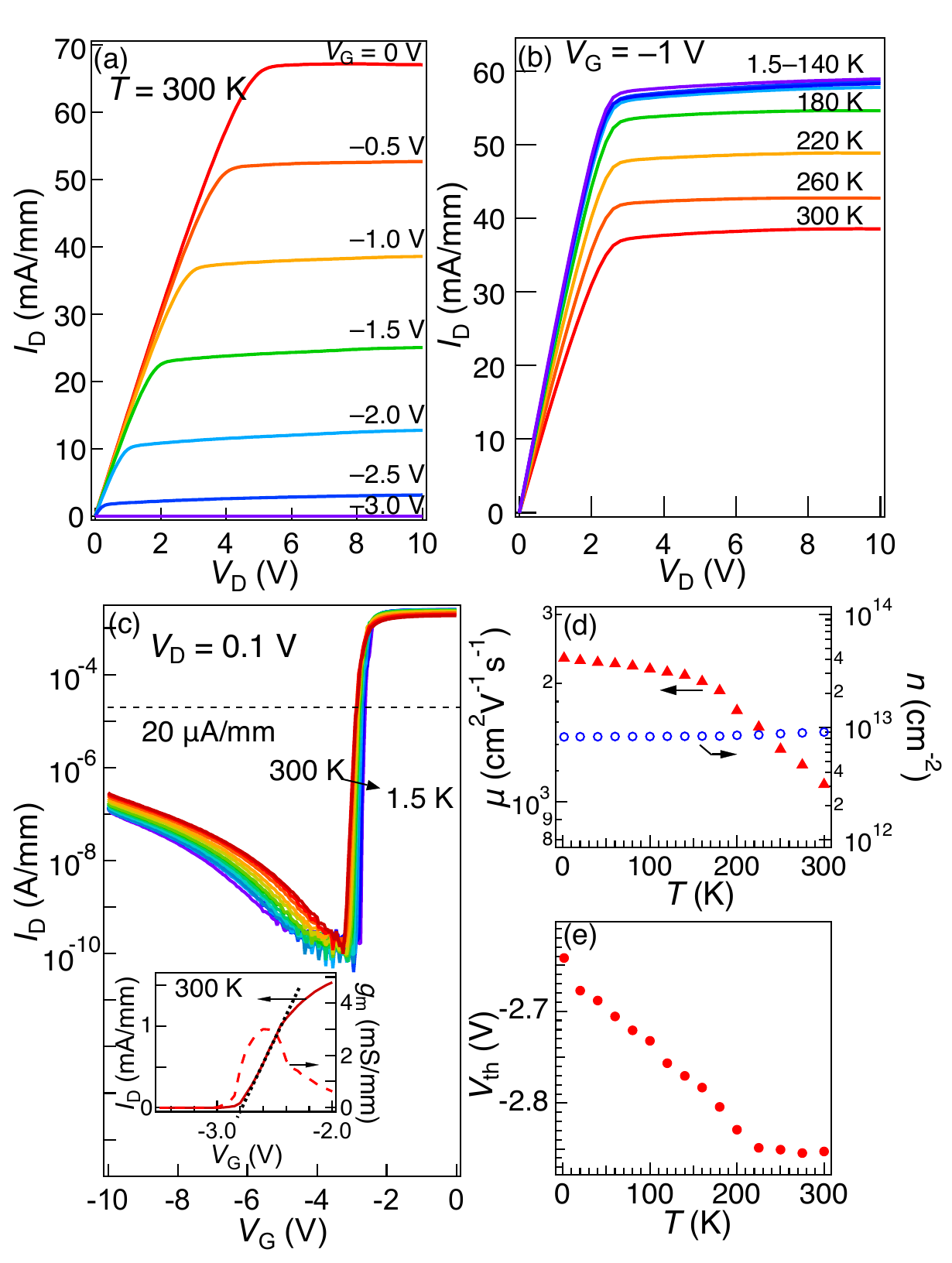}
\caption{\label{fig:IV}
(a) $V_\G$ dependence at $T=300$ K and (b) temperature dependence at $V_\G=-1$ V of $I_\D$ vs. $V_\D$. Curves below 180 K are taken at 140, 100, 60, and 1.5 K, respectively.  (c) Temperature dependence of $I_\D$ vs. $V_\G$ from 300 to 1.5 K (in steps of 25 K above 200 K, and in steps of 20 K below 200 K) at $V_\D = 0.1$ V. Inset: $I_\D$ vs. $V_\G$ at 300 K in the linear scale, and the corresponding transconductance $g_\mathrm{m}$. The dotted line is the linear extrapolation to extract $V_\mathrm{th}$.  (d) Mobility $\mu$ (triangles) and sheet electron density $n$ (hollow circles) of the ungated device as functions of temperature. (e) $V_\text{th}$ (extracted at $I_\D$ = 20 $\upmu$A/mm, the dashed line shown in (c)) as a function of temperature. }
\end{figure}

The electrical characteristics of the GaN HEMT are shown in Fig.~\ref{fig:IV}. Figs.~\ref{fig:IV}(a) and \ref{fig:IV}(b), respectively, display the gate-voltage ($V_\G$) dependence at room temperature and the temperature ($T$) dependence at $V_\G=-1$ V of the output characteristics (drain current $I_\D$ vs. drain voltage $V_\D$). It can be seen from Fig.~\ref{fig:IV}(b) that the channel conductivity increases as the temperature goes down, but its increase rate becomes very small below approximately 140 K. The temperature dependence of the transfer characteristics is shown in Fig.~\ref{fig:IV}(c), measured at a very small $V_\D$ of 0.1 V to minimize the influence of $V_\D$ for threshold voltage  ($V_\text{th}$) extraction. $V_\text{th}$ extracted at room temperature using the linear-extrapolation method (by locating the $V_\G$ intercept of the linear extrapolation of $I_\D$--$V_\G$ curve at its point of maximum transconductance, as shown in the inset of Fig.~\ref{fig:IV}(c), and then subtracting $V_\D$/2 \cite{Schroder2005,Tarasewicz1988}) is $-2.85$ V, which corresponds to an $I_\D$ of 20 $\upmu$A/mm. 

Fig.~\ref{fig:IV}(d) presents the low-field channel mobility of the ungated device,  extracted using a dc method based on the linear $I_\D$--$V_\G$ approximation \cite{Pradeep2016,Pradeep2017}, along with the sheet electron density determined by integrating the capacitance--$V_\G$ characteristics \cite{Schroder2005,Zhao2007}, with both quantities plotted as functions of temperature. The electron density is around $9 \times 10^{12}$ cm$^{-2}$ with a very weak temperature dependence. This value is consistent with the polarization-based analytical model \cite{Sharbati2021} and aligns well with experimental reports for structures with similar Al composition and AlGaN thickness \cite{Tan2011, Chen2025,Yuan2016}. The electron mobility, on the other hand, is 1108 cm$^2$/(V s) at room temperature and increases to 2323 cm$^2$/(V s) as $T$ goes down to 1.5 K. The rate of increase in the electron mobility with decreasing temperature substantially declines below approximately 140 K, which explains the weakened temperature dependence of the channel conductivity below 140 K as seen in Fig.~\ref{fig:IV}(b), and is similar with the temperature dependence of the 2DEG mobility of GaN HEMTs in other literature  \cite{Radhakrishnan2010, Hofmann2012, White2018}. The outstanding cryogenic performance of GaN HEMTs is primarily attributed to the nearly temperature-independent electron density and the increased mobility at lower temperatures. As the temperature decreases, acoustic phonon scattering is suppressed \cite{Sze2006}, leading to enhanced mobility. In contrast, Coulomb scattering from ionized impurities becomes more significant at lower thermal velocities, typically reducing mobility at lower temperatures. Nevertheless, since Coulomb scattering is weak in GaN \cite{Gui2020a}, electron mobility continues to increase at lower temperatures, although the rate of increase slows noticeably below  $\sim$140 K.

The dependence of $V_\text{th}$ on temperature, on the other hand, is displayed in Fig.~\ref{fig:IV}(e). The $V_\text{th}$ values here are extracted at $I_\D = 20$ $\upmu$A/mm using the constant-current method. $V_\text{th}$ shows negligible temperature dependence beyond 225 K, but then gradually shifts toward positive direction with decreasing temperature below 225 K. This may be explained by the frozen-trap effect: increased amounts of trapped electrons at lower temperatures due to inhibited detrapping process leads to the positive shift of $V_\text{th}$ \cite{Zeng2022,Chen2024,Tang2023}. The frozen-trap effect also strongly affects the gate leakage at cryogenic temperatures, as will be presented in the following paper.

\section{Temperature Dependence of the Gate Leakage}

\begin{figure}[!t]
\centering
\includegraphics[width=3.5in]{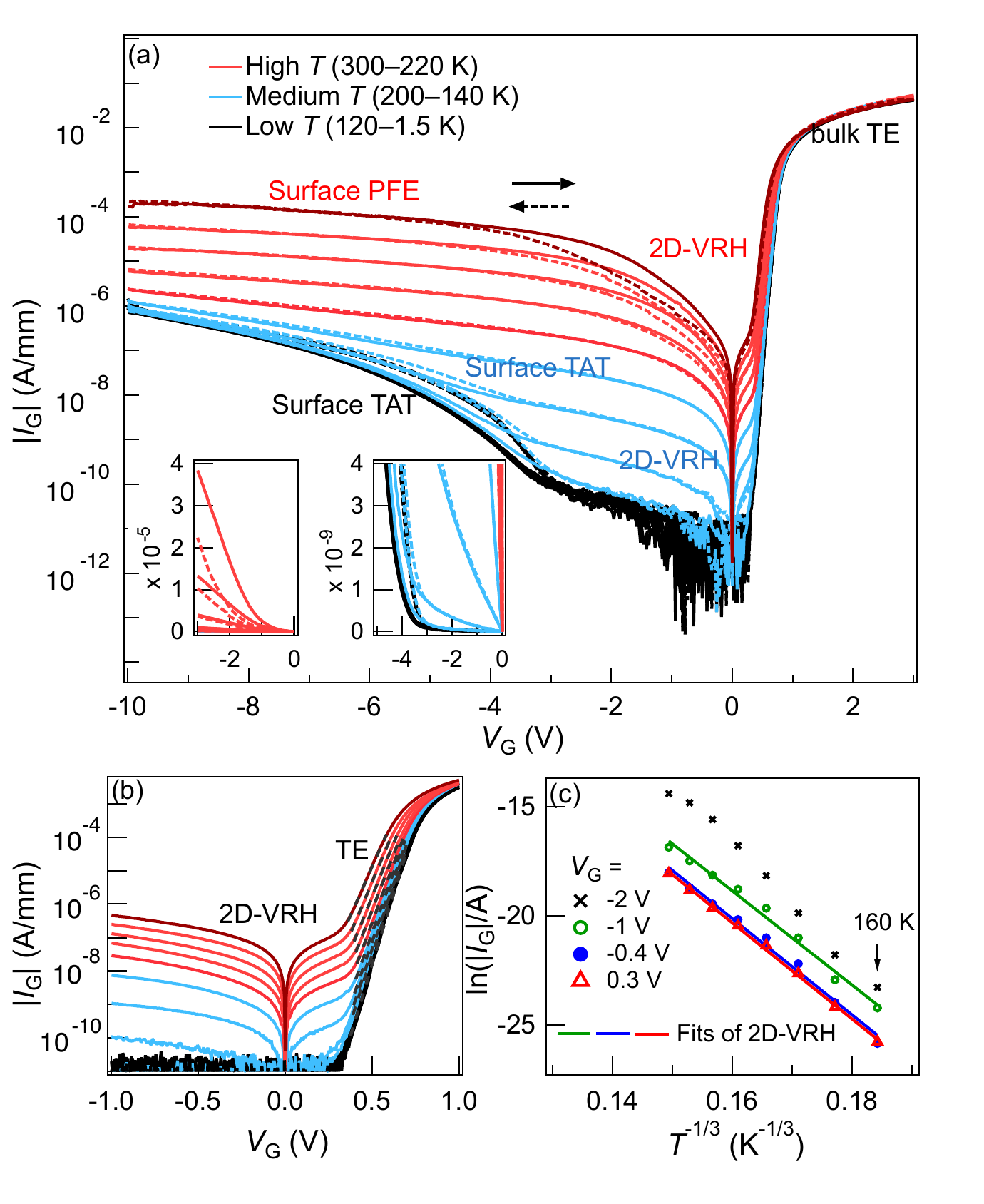}
\caption{\label{fig:IgVg}
(a) $I_\G$ vs. $V_\G$ from room to cryogenic temperatures in steps of 20 K. (The 300-K curve is specifically shaded in a darker red to distinguish the hysteresis loop.) The solid curves represent gate voltage sweeps from negative to  positive $V_\G$, and the dashed ones represent positive to negative $V_\G$ sweeps. Insets: same data in the linear scale. (b) Enlarged view of (a) for upward gate sweeps from $V_\G=-1$ V to $1$ V. Dashed lines are fits of the Schottky TE model. (c) Fits of the 2D-VRH model (Eq.~\ref{eq:2DVRH}) to $\ln(|I_\G|)$ vs. $T^{-1/3}$ at small $V_\G$'s $0.3$ V, $-0.4$ V, $-1$ V, and $-2$ V, respectively. $I_\G$ for $T<160$ K is too small with high noise-to-signal ratios and thus cannot be fitted.
}
\end{figure}

The source and the drain of the device were grounded for all gate-leakage measurements presented in the following figures. Fig.~\ref{fig:IgVg} shows the gate leakage current $I_\G$ vs. $V_\G$. The electrical characteristics can be classified into three temperature ($T$) categories \footnote{The temperatures indicated in this work are device temperatures instead of electron temperatures. The electron temperature can be higher than the device temperature, and may be extracted from temperature-dependent measurements of the Shubnikov-de Haas oscillations.}: high $T$ (300–220 K), medium $T$ (200–140 K), and low $T$ (120–1.5 K), with the low-$T$ category exhibiting negligible temperature dependence (i.e., all data collapse into the black curve). The solid curves represent \textit{upward} gate voltage sweeps from negative to positive $V_\G$, and the dashed ones represent \textit{downward} sweeps. An enlarged view for $-1$ V $\le V_\G \le 1$ V is displayed in Fig.~\ref{fig:IgVg}(b). For 0.3 V $\lesssim V_\G \lesssim 0.7$ V, the $I_\G$--$V_\G$ curves suit the TE model \cite{Sze2006}. For $V_\G \lesssim 0.3$ V, by contrast, the leakage $I_\G$ stops being exponentially suppressed by the Schottky barrier. Besides, unlike bulk leakage, which tends to saturate under reverse bias beyond the threshold voltage (i.e., $V_\G < V_\text{th}$) and hence results in a kink on the $I_\G$ curve at $V_\text{th}$ \cite{Turuvekere2013,Turuvekere2015}, $I_\G$ of our device keeps rising smoothly when passing $V_\text{th}$. To look into the gate-leakage mechanism in this regime, we consider possible \textit{surface} leakage current between the gate and the source and drain using the 2D-VRH model \cite{Xu2018,Kotani2007,Liu2011,Chen2014}: 
\begin{equation}\label{eq:2DVRH}
I_{\mathrm{2D}\text{-}\mathrm{VRH}} = I_0\exp \left[-\left(\frac{T_0}{T}\right)^{1/3}\right].
\end{equation}
The 2D-VRH model fits well to the leakage current at small $V_\G$'s from $+0.3$ V to $-1$ V for $T \ge 160$ K, which is presented as $\ln(|I_\G|)$ vs. $T^{-1/3}$ in Fig.~\ref{fig:IgVg}(c). Gate leakage at higher negative bias ($V_\G <-1$ V) starts to deviate from the 2D-VRH model to transition to other mechanisms, with an example of  $V_\G = -2$ V also shown in Fig.~\ref{fig:IgVg}(c). Gate leakage for $T<160$ K under small gate bias, on the other hand, is too small with high noise-to-signal ratios and thus cannot be analyzed. However, the mechanism transition is still evident in the electric-field dependence, which is especially clear the linear-scale plots of $I_\G$ vs. $V_\G$, as displayed in the insets of Fig.~\ref{fig:IgVg}(a). $I_\G$ begins to increase sharply around $V_\G = -1$ V at high $T$, whereas this onset shifts to approximately $V_\G =-3$ V for $T \le 160$ K. This may imply that, at low $T$, 2D-VRH may be dominant around zero gate bias and down to $V_\G =-3$ V, although the low-frequency noise of 2D-VRH is expected to exponentially grows with decreasing temperature \cite{Shklovskii2003}. 

\begin{figure}[!t]
\centering
\includegraphics[width=3.5in]{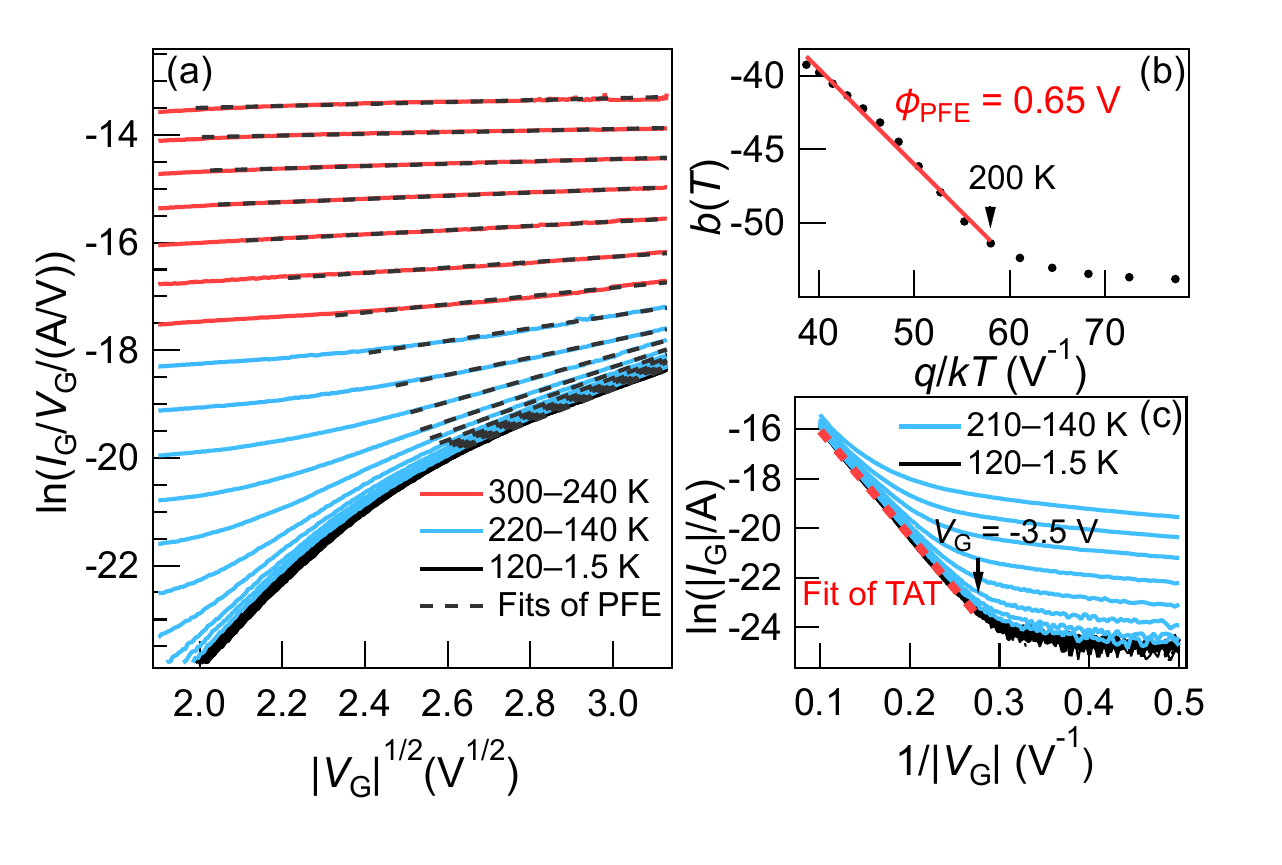}
\caption{\label{fig:PFEnTAT}
Gate leakage under higher negative $V_\G$ ($\lesssim-4$ V) for upward gate sweeps. (a) $\ln(I_\G/V_\G)$ vs. $\sqrt{|V_\G|}$ from room temperature to 1.5 K in steps of 10 K. Dashed lines are fits of the PFE model (Eq.~\ref{eq:PFE}). Parameters $b(T)$ from the fits are further plotted in (b) to obtain $\phi_\mathrm{PFE}$. (c) $\ln(|I_\G|)$ vs. $1/|V_\G|$ for 210--1.5 K. The dashed line is the fit using the TAT model (Eq.~\ref{eq:TAT}).
}
\end{figure}

More intriguingly, $I_\G$ under opposite directions of $V_\G$ sweeps can deviate from each other, as seen in Fig~\ref{fig:IgVg}(a). To investigate  the temperature dependence and the hysteresis behavior of the gate leakage, two main mechanisms of \textit{trap-assisted} surface leakage current are taken into account for higher negative $V_\G$: the Poole--Frenkel emission (PFE) \cite{Sze2006},  and the trap-assisted tunneling (TAT) \cite{Houng1999}, with the theoretical equations for the respective models listed below,
\begin{equation}\label{eq:PFE}
I_{\mathrm{PFE}} \propto E \exp \left[\frac{q}{k T}\left(\sqrt{\frac{q E}{\pi \epsilon}}-\phi_{\mathrm{PFE}}\right)\right],
\end{equation}
and 
\begin{equation}\label{eq:TAT}
I_{\mathrm{TAT}}\propto \exp \left(-\frac{8 \pi \sqrt{2 m^* q}}{3 h} \frac{\phi_{\mathrm{TAT}}^{3 / 2}}{E}\right),
\end{equation}
where $E$ is the electric field between the gate and the source and drain and is proportional to $V_\G$ within moderate gate bias \cite{Cao2021}, $q$ the elementary charge, $k$ the Boltzmann constant, $\epsilon$ the permittivity, $h$ the Planck constant, $\phi_{\mathrm{PFE}}$ and $\phi_{\mathrm{TAT}} $ the respective barrier heights of the trap states, and $m^*$ the electron effective mass, of which the weak temperature dependence gradually diminishes to be negligible as $T$ is decreased to approximately 100 K \cite{Hofmann2012,Korotyeyev2022,Armakavicius2024,Hubner2009,Sarkar2009,Barber1967}. $I_{\mathrm{PFE}}$ is exponentially enhanced with $T$, whereas $I_{\mathrm{TAT}}$ is nearly temperature independent at low temperatures, which is consistent with the curves below 140 K in Fig.~\ref{fig:IgVg}(a)  being completely overlapping. Both equations are exponentially related to $E$ and hence are more prominent at higher negative $V_\G$. Gate leakage for $V_\G \lesssim -4$ V is shown in Fig~\ref{fig:PFEnTAT}. Eq.~\ref{eq:PFE} can be written as $\ln(I_\G/V_\G)=m(T)\sqrt{|V_\G|}+b(T)$, where $b(T)=-q\phi_{\mathrm{PFE}}/kT + \ln C$ ($C$ is a constant). Attempted fits of the PFE model are presented in Fig.~\ref{fig:PFEnTAT}(a) as $\ln(I_\G/V_\G)$ vs. $\sqrt{|V_\G|}$, with the values of $b(T)$ from the linear fits (dashed lines in Fig~\ref{fig:PFEnTAT}(a)) further plotted in Fig.~\ref{fig:PFEnTAT}(b) as a function of $q/KT$. $\phi_{\mathrm{PFE}}$ can be obtained from the slope of Fig.~\ref{fig:PFEnTAT}(b) above 200 K, which is 0.65 V, and is close to the trap barrier heights observed from similar GaN-based HEMTs \cite{Zou2023}. However, the leakage current for $T < 200$ K deviates from the PFE model because of the frozen traps \cite{Wang2023, Jiang2023}. The TAT current starts to contribute for $T \lesssim 200$ K, and becomes dominant for $T \lesssim 120$ K, leading to data points that are nearly temperature independent in Fig.~\ref{fig:PFEnTAT}(b) at lower temperatures. Eq.~\ref{eq:TAT} fits well below 120 K for $V_\G<-3.5$ V, as presented in Fig~\ref{fig:PFEnTAT}(c) with the dashed line. $\phi_{\mathrm{TAT}}$ is proportional to the slope of the linear fit and can be determined if the relation between $V_\G$ and $E$ is  precisely known. However, according to simulations of the electric field near a GaN HEMT surface \cite{Liu2020,Cao2021}, $E$ is not uniform, peaking particularly around the electrodes, which predominantly determines the overall leakage magnitude. Although $\phi_{\mathrm{TAT}}$ may not be obtained easily from Fig.~\ref{fig:PFEnTAT}(c), its value is expected to match $\phi_{\mathrm{PFE}}$ in the same system.

\section{Temperature-Dependent Hysteresis of the Gate Leakage}

\begin{figure}[!t]
\centering
\includegraphics[width=3.7in]{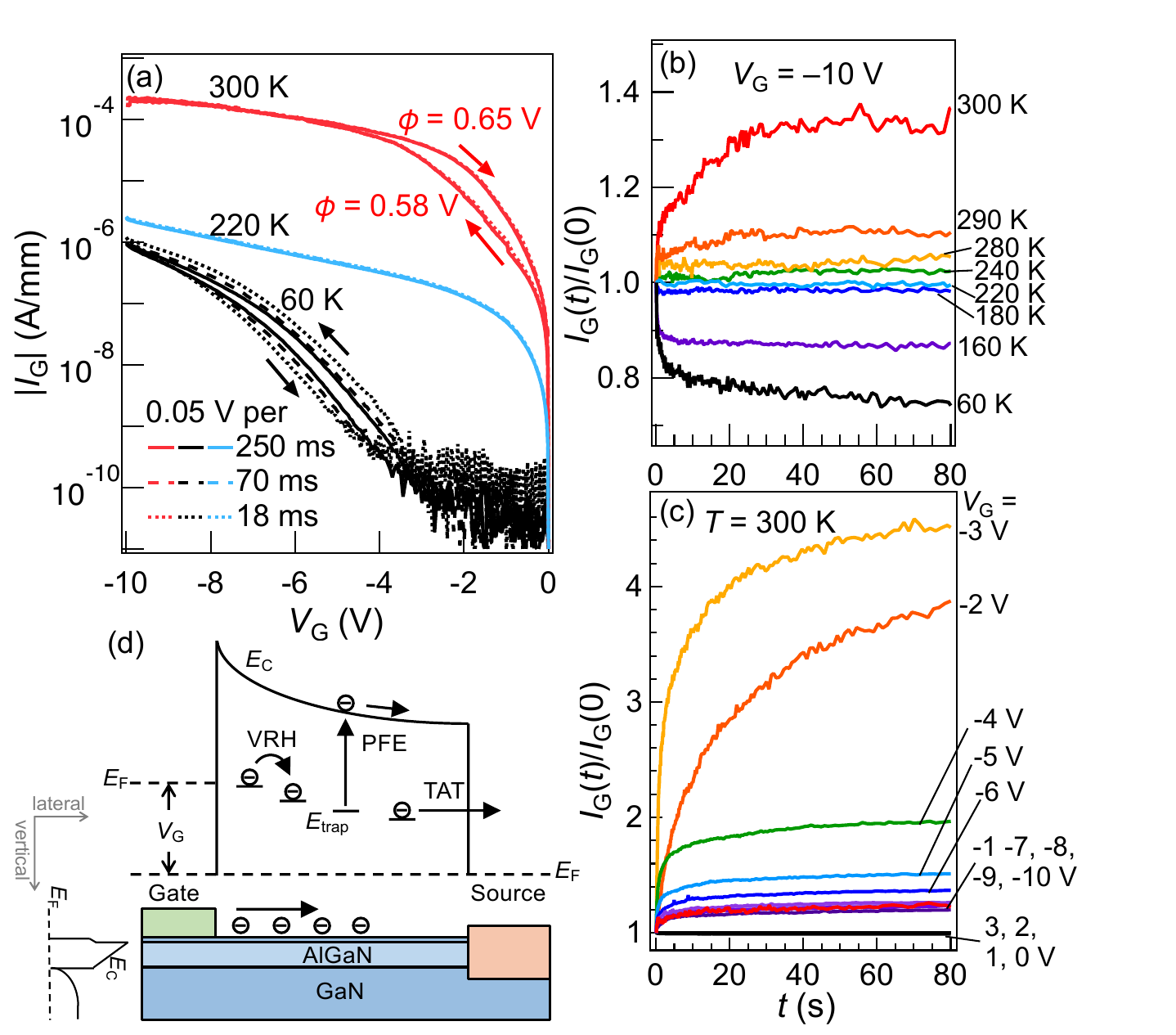}
\caption{\label{fig:IgVgh}
(a) $I_\G$ vs. $V_\G$ for both upward and downward gate sweeps. Data points are taken at a step size of $\pm0.05$ V per 250, 70, and 18 ms, respectively. $I_\G$ normalized to its initial quantity is plotted against stress time $t$ (b) under $V_\G=-10$ V at different temperatures and (c) at 300 K under different $V_\G$'s from 3 V to $-10$ V in steps of 1 V. (d) Schematic  diagram illustrating 2D-VRH, PFE, and TAT processes.
}
\end{figure}

The most captivating feature in Fig.~\ref{fig:IgVg} is the hysteresis loops of $I_\G$ under $V_\G$ sweeps, which even exhibit opposite loop directions at different temperatures. The loops are clockwise at higher temperatures (300–260 K) and counterclockwise at lower temperatures (200–1.5 K). The hysteresis is least obvious at the crossover temperatures (240–220 K) for our devices. Hysteresis loops at representative temperatures are redisplayed in Fig.~\ref{fig:IgVgh}(a) with various gate sweep rates from $\pm0.05$ V per 250 ms to $\pm0.05$ V per 18 ms. It can be seen that the shape of the loop is insensitive to the gate sweep rate at higher $T$ (unless the sweep rate is lowered down to $\pm0.05$ V per several seconds  to approach the saturation shown later in Fig.~\ref{fig:IgVgh}(b) to substantially minimize the hysteresis). For $T =60$ K, however, the hysteresis is highly correlated with the gate sweep rate; as the sweep rate increases, the hysteresis becomes more pronounced. The strong dependence of the hysteresis on the gate sweep rate is observed for all measurements at $T \lesssim 200$ K. 

To cross-check the history or time dependence of the leakage current, the shift in $I_\G$ normalized to its initial quantity ($I_\G(t)/I_\G(0)$) is also examined under a certain stress time $t$ with $V_\G=-10$ V at various temperatures, as shown in Fig.~\ref{fig:IgVgh}(b). For $T< 220$ K, $I_\G$ decreases over time until it approximately reaches a minimum. For $T> 220$ K, on the other hand, $I_\G$ increases over time until it reaches a saturation value. The change of the stress-time dependence of $I_\G$ from a decreasing trend to an increasing one was interpreted with hole injections in studies of $p$-GaN HEMTs at room temperature \cite{Chao2023, Shi2018}, which can not explain our case with undoped GaN. 220 K is also the crossover temperature around which the hysteresis loop of $I_\G$ vs. $V_\G$ reverses its direction during our temperature-dependent measurements. 

$I_\G(t)/I_\G(0)$ also exhibits $V_\G$ dependence, which reveals the dynamic charging process of the traps while the Fermi level varies. An example of $T=300$ K at various $V_\G$'s from 3 V to $-10$ V is shown in Fig.~\ref{fig:IgVgh}(c). For positive $V_\G$, where the leakage is dominated by TE, $I_\G$ shows no dependence on the stress time because it is not induced by traps. However, the time dependence emerges under negative $V_\G$. The rate of change in $I_\G$ over stress time becomes most pronounced at $V_\G= -3$ V, which corresponds to the widest part (on the linear scale, not shown) of the hysteresis loop being located around $V_\G= -3$ V. According to a simulation model of the gate leakage induced by surface traps \cite{Trew2009}, the $V_\G$ dependence indicates different surface trap densities at different energy levels with respect to the gate electrode; a steeper change in $I_\G(t)/I_\G(0)$ over time points to a higher surface trap density.

Interpretation of the hysteresis aids in gaining a deeper insight into the underlying mechanism of the surface leakage current. Fig.~\ref{fig:IgVgh}(d) illustrates the 2D-VRH, PFE, and TAT processes. With 2D-VRH at small $V_\G$, electrons hop between traps over varying distances. Under higher negative $V_\G$, PFE or TAT takes place with the assistance of electric field to lower the barrier. With PFE at higher temperatures, electrons are thermionically emitted from the traps into the conduction band, whereas with TAT at lower temperatures, electrons are frozen into the traps \cite{Wang2023, Jiang2023} and can only be emitted through tunneling. For $T \lesssim 200$ K, where the gate leakage is contributed by the surface TAT, a negative $V_\G$ would promote trap occupation by electrons \cite{Nicholls2019}, increasing the barrier height and thus suppressing the tunneling current. This results in a decrease of $I_\G$ under a stress time at a fixed $V_\G$ until all traps are occupied, or a counterclockwise hysteresis loop (as observed in Fig.~\ref{fig:IgVgh}(a)) when a high trap occupancy caused by a high negative $V_\G$ suppresses the $I_\G$ of a following upward sweep to a lower negative $V_\G$. A fit of Eq.~\ref{eq:TAT} to the downward gate sweep with a sweep rate of $\pm0.05$ V per 250 ms yields a $\phi_\mathrm{TAT}$ that is 11\% smaller than that of the upward sweep, which aligns with our interpretation. At higher temperatures where the surface PFE starts to take over, $\phi_\mathrm{PFE}$ obtained from the downward sweep is 0.58 V, which is also 11\% smaller than that of the upward sweep ($\phi_\mathrm{PFE}=0.65$ V from Fig.~\ref{fig:PFEnTAT}(b)) owing to the same reason. However, more electrons in the traps due to the stress of $V_\G$ does not point to an impediment of the PFE current but a promotion instead, because PFE electrons are liberated into the conduction band from the traps rather than tunneling between the traps. More electrons in the traps serve as a larger carrier reservoir that is ready to contribute to the leakage, leading to an increase of $I_\G$ under a stress time at a fixed $V_\G$, or a clockwise loop (in Fig.~\ref{fig:IgVgh}(a)) when a high negative $V_\G$ has prepared a large electron reservoir to later contribute to $I_\G$ in a following upward sweep to a lower $V_\G$. The increase of the surface PFE current due to the reservoir of trapped electrons can seriously backfire as surface effects become more pronounced in nanoscale modern devices. As for the pronounced reliance on the gate sweep rate for $T \lesssim 200$ K, this phenomenon is consistent with the frozen-trap effect, as the time needed for carrier trapping and de-trapping via tunneling overwhelms the gate sweep time. This can seriously affect the high-speed performance of cryogenic devices. 

The hysteresis behaviors are closely related to the detailed mechanisms of the surface leakage, serving as a quick and precise indicator of PFE or TAT depending on the direction of the hysteresis loop. GaN-based devices with other passivation materials (e.g., SiN \cite{Chen2014, Xu2018,Geng2018}, SiN/Al$_2$O$_3$ \cite{Liu2011}, or SiON \cite{Geng2018}) are also found to have trap-assisted surface leakage. SiO$_2$ passivation, while minimizing the shallow-trap density, tends to exhibit a higher density of deep traps relative to SiN or SiON \cite{Geng2018}. These surface traps are believed to originate from minor oxidation of the AlGaN barrier during the deposition of the SiO$_2$ layer \cite{Chevtchenko2007, Yagi2008}. While shallow traps are primarily responsible for high gate leakage, higher densities of deep traps are associated with pronounced hysteresis behavior. 

The fabrication method, on the other hand, also plays a critical role in determining device quality. 2D-VRH has been attributed to localized states that arise in disorder-rich regions or along dislocations \cite{Rackauskas2018}. Such defects commonly form during MOCVD growth due to lattice and thermal mismatch strain, high-temperature processing that promotes the formation of point and threading dislocations, surface dangling bonds at abrupt heterointerfaces, and unintentional impurity incorporation \cite{Hite2023,Hwang1997}. Defects generated during MOCVD, when located in the insulating dielectric layer, can also contribute to PFE and TAT under higher electric fields. In contrast, molecular beam epitaxy enables lower process temperatures and smoother surface morphology \cite{Ganguly2014}, which may help reduce surface traps \cite{Lin2004,Fariza2017,Bridger1999}. However, MOCVD remains dominant industrially due to its scalability and throughput, and leakage may be mitigated with proper passivation and gate stack engineering. The hysteresis measurements are reproducible across multiple gate-sweep cycles and even over different thermal cooldowns, making them a reliable and convenient diagnostic tool for identifying leakage mechanisms in future devices fabricated using improved modern processes.

\section{Conclusion}
This study offers a comprehensive examination of the gate leakage of GaN HEMTs from 300 K to 1.5 K. The observed leakage mechanisms include TE under positive gate bias, 2D-VRH near zero gate bias, and a transition from the surface PFE to the surface TAT around 220 K under negative gate bias. The trap barrier height extracted from the gate leakage current under the upward gate sweep is 0.65 V, which is 12\% higher than the 0.58 V obtained from the downward sweep likely due to higher trap occupancy under more negative gate bias. The transition from PFE to TAT around 220 K due to the frozen-trap effect induces significant opposite hysteresis in the leakage current under gate sweeps, exhibiting clockwise hysteresis loops above 220 K but counterclockwise ones below 220 K. The hysteresis below 220 K can be amplified under faster gate operation. The opposite hysteresis is explained by the differing emission mechanisms: PFE electrons are thermionically emitted from the traps into the conduction band, whereas TAT electrons are emitted via tunneling between traps. The hysteresis measurements can serve as a convenient tool to identify the leakage mechanisms. More experimental and theoretical research for the surface-trap-induced gate leakage are urgently required to help minimize the adverse effects of surface traps.

\section*{ACKNOWLEDGMENTS}
The work was supported by the National Science and Technology Council of the Republic of China under Grants No. NSTC 112-2112-M-003-018 and No. NSTC 113-2112-M-003-011.

%\section*{REFERENCES}
\bibliography{GaN}
\bibliographystyle{apsrev4-1-title.bst}

%\newpage

\end{document}